\documentclass[%
reprint, 
superscriptaddress,
 amsmath,amssymb,
 aps, prab
]{revtex4-2}

\usepackage{graphicx}
\usepackage{dcolumn}
\usepackage{bm}
\usepackage[T1]{fontenc}

\usepackage{amsfonts}
\usepackage{amsmath}
\usepackage{amssymb}
\usepackage{amsthm}
\usepackage{upgreek}

\usepackage{tikz}
\usetikzlibrary{arrows.meta}
\usetikzlibrary{calc}

\usepackage{float}
\graphicspath{{./images/}}

\usepackage[caption=false]{subfig}

\usepackage{empheq}

\usepackage{longtable}
\usepackage{array}
\usepackage{multirow}
\usepackage{adjustbox}

\usepackage[utf8]{inputenc}
\usepackage{mathtools}
\mathtoolsset{showonlyrefs}

\definecolor{purple}{rgb}{0.7,0,0.9}
\definecolor{lila}{RGB}{190, 80, 150}
\definecolor{blgr}{RGB}{0, 158, 115}
\definecolor{rod}{RGB}{225, 10, 0}

\begin{document}


\title{An Analytical Model for Fast Design Optimization and Performance Studies of nuSTORM}



\author{T. W. Choi}
\email[]{ting.choi@physics.uu.se}
\affiliation{Department of Physics and Astronomy, FREIA Division, Uppsala University, Uppsala, Sweden}

\author{M. Olvegård}
\email[]{maja.olvegard@physics.uu.se}
\affiliation{Department of Physics and Astronomy, FREIA Division, Uppsala University, Uppsala, Sweden}


\author{J. Aguilar}
\affiliation{Consorcio ESS-bilbao, Derio, Bizkaia, Spain}

\author{D. Barčot}
\affiliation{Center of Excellence for Advanced Materials and Sensing Devices, Ru\dj{}er Bo\v{s}kovi\'{c} Institute, Zagreb, Croatia}

\author{E. Baussan}
\affiliation{IPHC, Université de Strasbourg, CNRS/IN2P3, Strasbourg, France}

\author{A. K. Bhattacharyya}
\affiliation{European Spallation Source, Lund, Sweden}

\author{M. Blennow}
\affiliation{Department of Physics, School of Engineering Sciences, KTH Royal Institute of Technology, Stockholm, Sweden}
\affiliation{The Oskar Klein Centre, AlbaNova University Center, Stockholm, Sweden}

\author{M. Bogomilov}
\affiliation{Sofia University St. Kliment Ohridski, Faculty of Physics, Sofia, Bulgaria}

\author{B. Bolling}
\affiliation{European Spallation Source, Lund, Sweden}

\author{E. Bouquerel}
\affiliation{IPHC, Université de Strasbourg, CNRS/IN2P3, Strasbourg, France}

% ...

\collaboration{ESSnuSB Collaboration}
\email{ESSNUSB-DIS-L@in2p3.fr}
\noaffiliation


\date{\today}

\begin{abstract}
The concept of generating neutrino beams from stored muons (nuSTORM) has been explored for several decades. A few nuSTORM facilities have been designed for GeV muons, and a design of a sub-GeV nuSTORM, proposed for the European Spallation Source, is now underway. 
At low energy, the much larger beam divergence can significantly reduce the neutrino flux reaching the detector. Therefore, optimization of facility parameters is essential to avoid additional flux losses. This optimization, however, is not trivial due to the interdependency of the parameters. In this paper, we derive an analytical expression for the absolute neutrino flux incident on the detector front face and show how it can guide the optimization of a nuSTORM facility.
The analytical expression is furthermore validated with published data and Monte Carlo simulations from five separate cases. In addition, we show how it can be extended to also predict the neutrino energy distribution at the detector, thus offering a convenient method to estimate the facility performance without time-consuming Monte Carlo simulations of particle transport and decay. \end{abstract}


\maketitle

\section{Introduction}
One of the main purposes of a nuSTORM (neutrinos from STORed Muons) facility is the precise measurement of the neutrino cross-section. Conventional neutrino beams, for example those for long-baseline neutrino oscillation experiments, are produced by directing a high-intensity hadron beam onto a target, focusing the secondary pions produced in the process with a magnetic horn. The pions are then allowed to decay in an evacuated tunnel to produce neutrinos. Uncertainties in pion production and focusing, however, translate into sizable neutrino beam systematic errors, making such a scheme unsuitable for neutrino cross-section measurements \cite{nuflux1997}.

A different approach, nuSTORM, was proposed around 1980~\cite{nustorm1980} with neutrino beams produced from muon decay. The concept of nuSTORM is to store either $\mu^+$ or $\mu^-$ in a racetrack ring whose two long straight sections act as decay channels for muon decays through $\mu^+ \rightarrow e^+ + \nu_{e} + \bar{\nu}_{\mu}$ and $\mu^- \rightarrow e^- + \bar{\nu}_{e} + \nu_{\mu}\,.$

Unlike the pion beam in a decay tunnel, the muon beam intensity can be monitored to percent-level precision~\cite{nustorm_cern}, since it is stored over multiple turns, which makes the resulting flux of both electron neutrinos and muon neutrinos very well characterized. This enables absolute cross-section measurements and, in addition, other physics searches such as for sterile neutrinos.

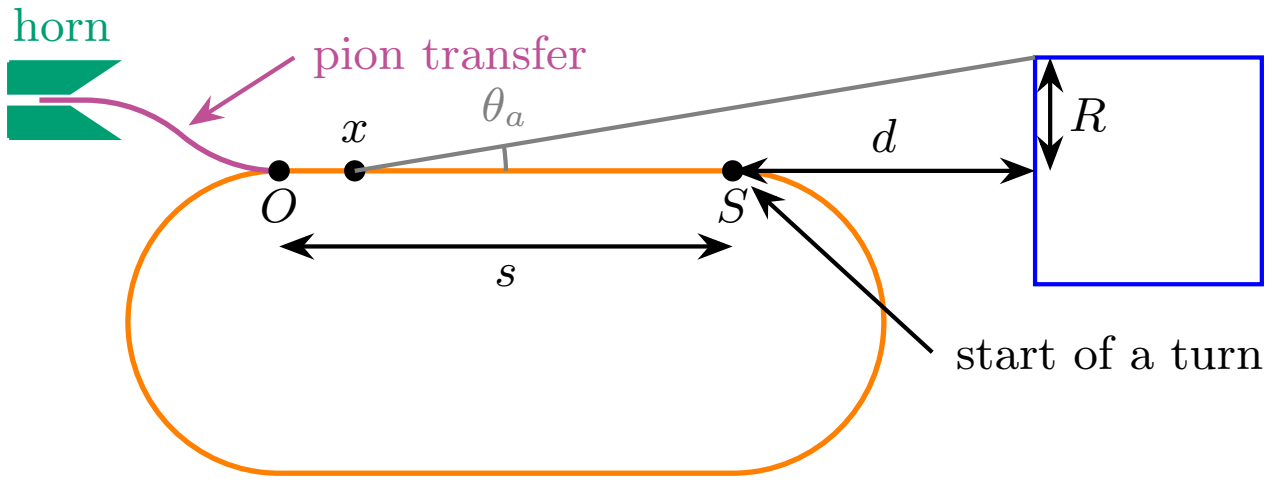
\begin{figure}[H]
  \centering
    \begin{tikzpicture}[>=Stealth,thick, every node/.style={font=\small}]
        \def\a{1}   
        \def\S{3.0}   
        \def\d{2.0}   
        \def\R{0.75}   
        \draw[orange,line width=1pt]
            (0,  \a)  arc (90:270:{\a})             
            (0,  \a) -- ++(\S,0)                      
            (0, -\a) -- ++(\S,0)                     
            arc (-90:90:{\a}) ;
        \draw[lila,line width=1pt] (0,  \a)  arc (-90:-130:{\a}) arc (-130:-90:{-\a}) coordinate (horn) --++(-0.3,0);
        \draw[lila, <-] (0,\a)++(-0.6,0.3) -- ++(0.7,0.45*\a) node[right,lila] {pion transfer};


        \filldraw[blgr] (horn)++(-0.5,-0.25) --++(0.7,0) --++(-0.3,0.2) --++(-0.4,0) --++(0,-0.2);
        \filldraw[blgr] (horn)++(-0.5,0.25) --++(0.7,0)--++(-0.3,-0.2) --++(-0.4,0) --++(0,+0.2);
        \node[blgr] at ($(horn)+(-0.15,0.5)$) (hornlabel) {horn};

        \coordinate (O)  at (0,  \a);
        \coordinate (s)  at (\S, \a);
        \coordinate (X)  at (0.5,\a);        
        \coordinate (OD)  at (\S+\d, \a);        
        \coordinate (ODt)  at (\S+\d, \a+\R);        
        \fill (O) circle(2pt) node[below] {$O$};
        \fill (X) circle(2pt) node[above=2pt] {$x$};
        \fill (s) circle(2pt) node[below] {$S$};
        \draw[<->] ([yshift=-5mm]O) -- node[below] {$s$} ([yshift=-5mm]s);
        \draw[black, <-] (s) ++(0.12,-0.1) -- ++(1.2,-1.1*\a) node[below, right=1pt,black] {start of a turn};
        
        \draw[blue] (OD) ++(0,\R) rectangle ++(2*\R,-2*\R);  
        \draw[<->] (s) -- node[above] {$d$} (OD);
        \draw[<->] ([xshift=1mm]OD) -- node[right] {$R$} ([xshift=1mm]ODt);
        \draw[gray] (X) -- (ODt) ++(-\R,0);           
        \draw[gray] (X)+(1,0) arc(0:10:1) node[pos=0.6,above=2pt] {$\theta_a$};
        
        \coordinate (ODe)  at (\S+\d+2*\R, \a);        
        \coordinate (ODe2)  at (\S+\d+3*\R, \a);        
    \end{tikzpicture}
    \caption{Plan‐view of the racetrack ring and the detector (blue square). For simplicity, the detector is assumed to be a right circular cylinder with a radius $R$. The central axis of the straight section is aligned with that of the detector.}
    \label{fig:racetrack}
\end{figure}

A nuSTORM facility is a short-baseline neutrino experiment that consists of a proton driver and a target, a magnetic horn, a pion-transfer beamline, a muon storage ring, and a neutrino detector, as illustrated in Fig.~\ref{fig:racetrack}. It starts with an intense proton beam that strikes a target embedded inside the horn to produce pions. The pions are then focused by the horn and directed into the pion transfer line. Pions will be injected into the racetrack storage ring at the start of one of the straight sections. As they propagate through the straight section, the pions decay and yield muons that span a broad momentum range, of which only the fraction within the transverse and momentum acceptance is captured and stored in the ring. Muons that fall outside of the acceptance, as well as any remaining pions, will be dumped at the end of the straight section or lost in the arc. 

At a certain distance away from the end of one of the straight sections, called the production straight, is the neutrino detector. The stored muons circulate until they decay or are lost on the aperture. A neutrino created in the production straight section can reach the downstream detector only if its emission angle at the creation point is smaller than the acceptance angle of the detector. A drawing of the facility geometry is shown in Fig~\ref{fig:racetrack}.

A complete nuSTORM design for GeV beams was developed in 2015 for Fermilab (Fermi National Accelerator Laboratory)~\cite{Geer2011, fermi_nu} and a modified design was developed for CERN~\cite{nustorm_cern}. Within the ESSnuSB+ (European Spallation Source neutrino Super Beam plus) project, the nuSTORM concept is extended to a Low-Energy nuSTORM (LEnuSTORM), optimised for the 0.2\,–\,0.6\,GeV neutrino energy range relevant to the ESSnuSB long-baseline experiment~\cite{ESSnuSB_CDR,essnusb2023}.

The performance of a nuSTORM facility depends on many design variables, such as the muon and pion momenta, the length the straight section, and the detector baseline. These must all be tuned to balance the neutrino yield and the facility cost, while considering other design constraints such as the energy of the proton driver. In this paper, we will focus on the geometry of nuSTORM, with particular emphasis on the length of the production straight section, $s$. 

Previous designs at Fermilab and CERN claim that the length $s$ has been chosen so that roughly 50\% of the injected pions decay in the straight section~\cite{fermi_nu}. However, that is not necessarily optimal. Making the straight section longer increases the number of pion decays, but also increases the risk that generated neutrinos miss the detector. This interplay becomes particularly pronounced in the low-energy case. 

Here, we derive an analytical expression that links the absolute neutrino flux at the detector front face to all geometric facility parameters. With reasonable assumptions for the other quantities, we can plot the expression as a function of $s$, revealing the approximate neutrino flux that can be expected from the facility and the optimal design parameters, before any detailed accelerator lattice design or Monte Carlo (MC) simulation is performed.

The expression can also be used to estimate the energy distribution of the neutrinos, which can be used for studying the performance of the cross-section measurement from an early point in the design of a nuSTORM facility.

In this paper, we will first describe the derivation of the analytical expression for the total neutrino flux at the detector. We will then show how it can be used to optimize the length of the straight section. We will compare the calculated number of neutrinos with equivalent estimates obtained from Monte Carlo simulations of full designs, and we will show similar comparisons of the neutrino energy distributions.

\section{Derivation}
We start with assuming a monochromatic pencil beam of pions injected at point $O$ in Fig.~\ref{fig:racetrack}. We imagine that $N_\pi$ pions enter the production straight and travel a distance $s$. Each pion decays with the time-dilated mean lifetime, $\gamma_\pi\tau_\pi$, which corresponds to the characteristic decay length
\begin{equation}
    L_{\pi} = \beta_\pi\, c\,\gamma_\pi \,\tau_\pi \,  ,
    \label{eq:decay_length}
\end{equation}
where $c$ is the speed of light, and $\beta_\pi$ and $\gamma_\pi$ are the relativistic speed and Lorentz factor, respectively. From the pion decays, the number of muons by point $S$ is 
\begin{equation}
    N_\pi (1-e^{-s/L_\pi})
    \label{eq:pion_decay}
\end{equation}
The ring admits muons whose momenta lie within a half-window momentum deviation, $\delta_{acc}$, about the muon reference momentum $P_\mu$. The muon momentum spectrum produced by a monochromatic pion beam is uniform; for the reasoning, please refer to Appendix~\ref{app:uniform_Pmu}. The fraction of captured muons, therefore, is
\begin{equation}
    F_{\updelta}= \frac{2\,P_\mu\,\delta_{acc}}{P_{\max} - P_{\min}},
    \label{eq:muon_mom_frac} 
\end{equation}
where $P_{\max} $ and $ P_{\min}$ are the maximum and minimum muon momenta, respectively, obtainable from the pion decay and are functions of pion momentum $P_\pi$. In the relativistic limit $\beta\approx1$,  $P_{\max} \simeq P_\pi$ and $P_{\min} \simeq 4P_\pi/7$, so that equation~\eqref{eq:muon_mom_frac} can be simplified to:
\begin{equation}
    \frac{2\,P_\mu\,\delta_{acc}}{P_{max} - P_{min}} \simeq \frac{14P_\mu\,\delta_{acc}}{3P_\pi}.
\end{equation}
For the full derivation, please refer to Appendix \ref{app:47ratio}. 

We assume in this derivation that the ring admits muons with all emission angles, angles which can be calculated from the two-body decay kinematics. A real beamline, however, may have a finite acceptance on the muon angle or divergence from the focusing and aperture choices. This must be considered through the choice of pion and muon momenta, in combination with a matching beamline design.

After one full revolution from $S$ to $S$, the fraction of muons that have decayed is 
\begin{equation}
    1-e^{-C/L_\mu}, 
    \label{eq:muons_decay_in_one_turn}
\end{equation}
where $C$ is the circumference of the ring and $L_\mu$ is the muon decay length, calculated using Eq.~\eqref{eq:decay_length} but with suffix $\pi$ replaced by $\mu$. If we assume that muons survive many turns in the ring, i.e. that $C \ll L_\mu$, the number of muon decays during each turn can be approximated as being uniformly distributed throughout the ring. The fraction $F_s$ of muons decaying in the production straight is then essentially the straight-to-circumference ratio. For a racetrack ring with an arc length $a$, that ratio is 
\begin{equation}
    F_s=\frac{s}{2(a+s)}=\frac{1}{2\left(1+a/s\right)} \,,
    \label{eq:s2c}
\end{equation}

The next aspect to consider is that a neutrino created at position $x$ along the straight section reaches the detector only if it is emitted within the acceptance angle of the detector. For simplicity, the detector is assumed to be a right circular cylinder with a radius $R$, and only the circular front face is considered in this derivation. We obtain an acceptance angle
\begin{equation}
    \theta_a(d+s-x) = \tan^{-1}\left(\frac{R}{d+s-x}\right) \, .
    \label{eq:theta_x}
\end{equation}
where $d+s-x$ is the distance from the detector front face to the point $x$ where the muon decays. 

In the lab frame, the angular distribution of neutrinos emitted from relativistic muons can be described by relativistic beaming and is given by 
\begin{equation}
    \label{eq:headlight}
    \eta(\theta)=\frac{1}{2}\sin\theta \,\frac{1-\beta^{2}}{\left(1-\beta\cos\theta\right)^{2}}\,d\theta. 
\end{equation}
The derivation of this expression is available in Appendix~\ref{app:headlight}. In Fig.~\ref{fig:headlight} we show it for 3.8\,GeV/c and 0.6\,GeV/c parent muons. These numbers correspond to the maximum muon momenta of the Fermilab and ESS-based nuSTORM facilities, respectively. For low-energy muon decay, the angular distribution is much wider, and the peak angle is shifted upwards, increasing the risk of neutrinos falling outside of the detector acceptance.
\begin{figure}[h]
    \centering
    \includegraphics[width=1\linewidth]{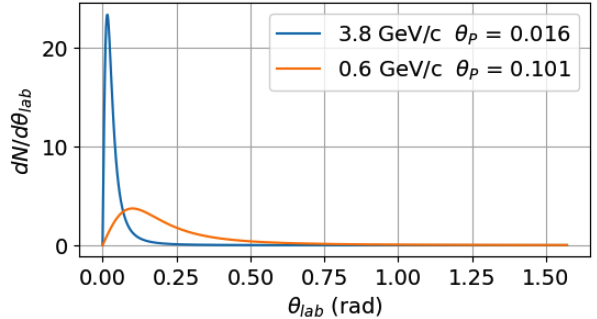}
    \caption{The angular density distribution of neutrinos emitted from 3.8\,GeV/c (blue) and 0.6\,GeV/c (orange) muons. The most probable emission angle for each case, $\theta_{\text{P}}$, is defined at the peak of the curve.}
    \label{fig:headlight}
\end{figure}

To get the neutrino acceptance probability, $F_D(d+s-x)$, i.e. the fraction of neutrinos that end up at the detector out of the total number of neutrinos produced in the muon decay, we integrate Eq.~\eqref{eq:headlight} from 0 to $\theta_a$ defined in Eq.~\eqref{eq:theta_x}:
\begin{equation}
    \begin{aligned}
    F_D(d+s-x) &=\int_{0}^{\theta_a}  
    \frac{1}{2}\sin\theta \, \frac{1-\beta^{2}}{\bigl(1-\beta\cos\theta\bigr)^{2}}
    \, d\theta \\
    &=  \frac{1}{2}   \frac{(1+\beta)(1-\cos\theta_a)}{1-\beta\cos\theta_a} \,.
    \end{aligned}
\end{equation}

Since the muon density along the straight section falls exponentially, more muons decay upstream than downstream. We normalize it to the length $s$ and obtain the muon density at position $x$, $F_\mu(x,\,s)$: 
\begin{equation}
    F_\mu(x;\,s)=
    \begin{cases}
    \dfrac{e^{-x/L_\mu}}
    {L_\mu\left(1-e^{-s/L_\mu}\right)}, & 0\le x\le s,\\[6pt]
    0, & \text{otherwise}.
    \end{cases}
    \label{eq:mu_den}
\end{equation}
$F_D$ and $F_\mu$ are then convoluted along the straight section as
\begin{gather}
    \left(F_D * F_\mu\right)(d+s)= \int_0^s F_\mu(x;\,s)F_D(d+s-x) \,dx= \nonumber\\
    \frac{ 1}{2L_\mu(1-e^{-s/L_\mu})}   \int_{0}^{s} e^{-x/L_\mu} \,
     \frac{(1+\beta)(1-\cos\theta_a)}{1-\beta\cos\theta_a} \, 
    dx.
    \label{eq:convol}
\end{gather}

Now we combine Eqs.~\eqref{eq:pion_decay}, \eqref{eq:muon_mom_frac}, \eqref{eq:muons_decay_in_one_turn}, \eqref{eq:s2c} and \eqref{eq:convol} into an expression for the neutrino yield at the detector from a single turn:
\begin{gather}
    N_\nu^{(1)} = N_\pi 
     F_{\updelta}
     \,
    \frac{(1-e^{-s/L_\pi})(1-e^{-C/L_\mu})}{4L_\mu\bigl(1-e^{-s/L_\mu}\bigr)\left(1+\dfrac{a}{s}\right)} \, \nonumber \\
    \times \int_{0}^{s} e^{-x/L_\mu} \,
    \frac{(1+\beta)(1-\cos\theta_a)}{1-\beta\cos\theta_a} \,  dx \,.
    \label{eq:Nnu_one}
\end{gather}


Because muons will complete many revolutions, the neutrino flux must be accumulated over successive turns. Before summing over successive turns, recall that the stored muon population decays from turn to turn. After each revolution of length $C$, the number of muons at the start of the next turn is multiplied by the factor $\exp(-C/L_\mu)$:
\begin{equation}
    \begin{aligned}
        T_0 &:\quad N_\nu^{(1)} \\
        T_1 &:\quad N_\nu^{(1)} e^{-C/L_\mu} \\
        T_2 &:\quad N_\nu^{(1)} e^{-2C/L_\mu} \\
        \vdots &\quad\quad\quad\vdots \\
        T_n &:\quad N_\nu^{(1)} e^{-nC/L_\mu} \, ,
    \end{aligned}
\end{equation}
To obtain the total flux, we sum the contribution from turn 0 to infinity. The sum is a geometric series with an analytical solution given by
\begin{equation}
    N_\nu =\sum_{n=0}^{\infty} N_\nu^{(1)} e^{-nC/L_\mu}  = \frac{N_\nu^{(1)}}{1-e^{-C/L_\mu}} \\
    \label{eq:Nnu_sum}
\end{equation}

We observe that the geometric summation cancels out the $1-e^{-C/L_\mu}$ factor in Eq.~\eqref{eq:Nnu_one}, meaning that all initial muons will have decayed, equivalently to the case of an infinitely long straight section.

Lastly, we introduce a ring efficiency factor, $\epsilon_{r}$ with $0<\epsilon_r\leq1$, which is a lattice-dependent factor that depends on the performance of the lattice resulting from, e.g., the choice of beam aperture, beam losses due to resonance, etc. After a lattice design has been made, multi-particle tracking studies can be used to make a precise estimate of the lattice efficiency parameter. Until such simulations are in place, the choice of $\epsilon_r$ is based on a reasonable estimate, with $\epsilon_r \approx 0.5$ expected to be a realistic value.

The total number of neutrinos per flavor at the detector is now
\begin{equation}
    \begin{aligned}
        N_\nu(s;\, a,\, d,\, R,\,P_\mu,\, P_\pi, \,N_\pi, \,\delta_{\rm{acc}}, \, \epsilon_r) 
        =
        \frac{7N_{\pi}\,\delta_{acc}\,\epsilon_{r}}{6}\frac{P_\mu\,}{P_{\pi}} \, \\
        \times
        \frac{\bigl(1-e^{-s/L_\pi}\bigr)}{L_\mu\bigl(1-e^{-s/L_\mu}\bigr)\left(1+\dfrac{a}{s}\right)} \,  \\
        \times \int_{0}^{s} e^{-x/L_\mu} \,
          \frac{(1+\beta)(1-\cos\theta_a)}{1-\beta\cos\theta_a} \,
        dx \,,
    \end{aligned}
    \label{eq:N_nu}
\end{equation}
where $\theta_a(d+s-x)$ is defined in Eq.~\eqref{eq:theta_x}. This expression lets us estimate the absolute neutrino flux at the detector based on geometry and beam-line parameters, together with the muon and pion momenta and the relative momentum acceptance of the ring. 

If the initial protons-on-target count $N_p$ and the combined efficiency of the horn and the pion transfer line, $\epsilon_{\pi}$, are known, we can also include the term $N_\pi$ expressed as
\begin{equation}
    N_\pi=N_p \, \epsilon_{\pi}\,.
\end{equation}

Because there is no simple analytical expression for the remaining integral in Eq.~\eqref{eq:N_nu}, we will in the next section compute it numerically in order to compare it with Monte Carlo simulations. 

\section{Validation \label{sec:valid}}
The expression in Eq.~\eqref{eq:N_nu} has been benchmarked with Monte Carlo simulations for five different cases, where we have compared the total number of electron neutrinos, $\nu_e$, crossing the detector front face. The calculated number $N_\nu$ is compared with the simulated number $N_{\nu,{\rm MC}}$ and the relative discrepancy is calculated as
\begin{equation}
   \Delta =  \frac{(N_\nu-N_{\nu,\,\text{MC}})}{N_{\nu,\,\text{MC}}}  \,.
\end{equation}

In Cases 1 and 2 we use parameters extracted from the nuSTORM design for Fermilab presented in 2015~\cite{fermi_nu}. This design is based on a FODO (focusing-defocusing) lattice with a 180\,m straight section, 59.8\,m arc, 5\,GeV/c reference pions and 3.8\,GeV/c reference muons. The design has a lattice efficiency factor of $\epsilon_r =0.67$, taken directly from the muon survival rate without considering decay, and a momentum acceptance of $\pm$0.1. The design is based on having two detectors at separate locations: at $d=50$\,m, which we label Case 1, and with the $d=2050$\,m baseline, which we call Case 2.

We extract the total flux $N_{\nu,\,\text{MC}}$ visually from Fig.~5.16 in Ref.~\cite{fermi_nu}. The original paper provides energy spectra for both neutrino flavors, but here we only analyse the $\nu_e$ fluxes. The paper expresses the neutrino flux normalized to surface area without specifying the detector size. For Case 1 we use $R=2.82$\,m, which has been approximated from the detector size of 5\,m$\times$5\,m, specified in Ref.~\cite{nustorm2022}. For Case 2 we use $R=3$\,m, which is found separately in Ref.~\cite{nu_fermi_far}. For ease of comparison of the neutrino flux, we normalize the calculated number of neutrinos crossing the front surface by the surface area.

The input parameters together with the flux values are listed in Table~\ref{tab:valid_data_fermi}, with the results normalized to $10^{21}$\, protons on target. We observe that Eq.~\eqref{eq:N_nu} can predict the neutrino flux within one order of magnitude without extensive lattice design efforts and time-consuming Monte Carlo simulations.
\begin{table}[htbp]
    \renewcommand{\arraystretch}{1.2}
    \centering
    \caption{Validation data with parameters based on the \mbox{nuSTORM} design for Fermilab~\cite{fermi_nu, nustorm2022, nu_fermi_far}, assuming $N_p=10^{21}$ protons on target.}
    \label{tab:valid_data_fermi}
    \begin{tabular}{c|ccccc}\hline
        Case            & 1 & 2 \\ \hline
        $d$ [m]         & 50   & 2050   \\
        $R$ [m]         & 2.82  & 3    \\
        $s$ [m]         & 180 & 180   \\
        $a$ [m]         & 59.8 & 59.8   \\
        $P_\pi$ [GeV/c] & 5 & 5      \\
        $P_\mu$ [GeV/c] & 3.8 & 3.8    \\
        $\epsilon_r$    & 0.67 & 0.67   \\
        $\delta_{acc}$  & 0.1 & 0.1   \\
        $N_p$           & $10^{21}$ & $10^{21}$ \\ 
        $N_{\nu,\,\text{MC}}$ [m$^{-2}$] \quad &  \quad $4.7\times10^{16}$ \quad & $5.8\times10^{14}$ \\
        $N_\nu$  [m$^{-2}$]              &  \quad$4.3\times10^{16}$  \quad& $2.5\times10^{14}$  \\
        $\Delta$ [\%]   & -7.5 & -57 \\ \hline
    \end{tabular}
\end{table}

Cases 3 to 5, shown in Table~\ref{tab:valid_data_ESS}, relate to the low-energy nuSTORM design made for the ESSnuSB+ project~\cite{essnusb2023}. The ring features 81\,m straight sections, 27\,m arcs, and a detector of 2.5\,m radius positioned 50\,m downstream from the first straight section. The ring is fully FODO-cell structured with a 0.2\,m aperture radius. For the detailed design, please refer to Ref.~\cite{choi:ipac2026-mop1098}. 
\begin{table}[htbp]
    \renewcommand{\arraystretch}{1.2}
    \centering
    \caption{Validation data with the LEnuSTORM design for the ESSnuSB project. The flux is computed from $N_\pi=6\mathrm{E}5$ primary pions.}
    \label{tab:valid_data_ESS}
    \begin{tabular}{c|ccccc}\hline
        Case             &  3 & 4 & 5 \\ \hline
        $d$ [m]          & 50 & 50 & 50   \\
        $R$ [m]          & 2.5 & 2.5 & 2.5  \\
        $s$ [m]          & 81 & 81 & 81    \\
        $a$ [m]          & 27 & 27 & 27   \\
        $P_\pi$ [GeV/c]  & 1   & 3   & 5    \\
        $P_\mu$ [GeV/c]  & 0.6 & 2.2 & 3    \\
        $\epsilon_r$     & 0.5 & 0.5 & 0.5  \\
        $\delta_{acc}$   & 0.1 & 0.1 & 0.1  \\
        $N_\pi$    & 6E5 & 6E5 & 6E5 \\ 
        $N_{\nu,\,\text{MC}}$ & 800 & 6400 & 4100  \\
        $N_\nu$               & 720 & 4100 & 3200 \\
        $\Delta$ [\%]    & -12 & -37 & -21 \\
        \hline
    \end{tabular}
\end{table}

We estimate the values of $\delta_{acc}$ and $\epsilon_r$ through particle tracking simulations with Xsuite~\cite{xsuite}. We used a water-bag distribution in phase space with 99\% beam emittance of 8.9\,mmrad, which is the linear transverse acceptance of the ring. The momentum acceptance of the ring is evaluated by first assigning a random, uniformly distributed relative momentum deviation in the range from -0.1 to 0.1 to the water-bag distribution. The beam is then tracked in the ring for 40 turns, after which 90\% of the muons would have decayed (although decay is not activated in the tracking simulation). The surviving particles are registered after the tracking, and $\delta_{acc}$ is extracted as the energy range of the surviving muons. The ring efficiency $\epsilon_r$ is the muon survival rate, again without considering decay.

The ring design was migrated to BDSIM~\cite{BDSIM} for Monte Carlo simulations of the pion, muon and neutrino propagation. A monochromatic pencil beam of $6\times10^5$ primary pions was launched at the injection point of the ring. The pions and daughter muons were then tracked through the ring with particle decay activated. A 0.2\,m global aperture is set, which then kills the particles that cross it. The six-dimensional phase space coordinates of neutrinos were registered at their creation. The coordinate at the detector front face is calculated afterwards. From that, the total neutrino flux was calculated once the detector radius is fixed. 

The results of Case 3~-~5 are listed in Table~\ref{tab:valid_data_ESS}. We observe that the numerical calculation consistently underestimates the flux compared to the Monte Carlo simulation. The reasons are not fully understood. One of the reasons is that $\epsilon_r$ is underestimated because muons decay in reality.

 Nevertheless, all five datasets, Case 1~-~5, give discrepancies $\lvert \Delta \rvert <100\%$. In other words, Eq.~\eqref{eq:N_nu} predicts the neutrino flux to the correct order of magnitude across widely separated momentum regimes, making it a practical tool for rapid studies during the early phases of a facility design.
\section{Optimizing the straight length}
The validation in Sec.~\ref{sec:valid} confirms that the analytical expression predicts the absolute neutrino flux with order-of-magnitude accuracy. Other ways of use are possible. In this section, we will show using Eq.\eqref{eq:N_nu} for facility geometry optimization.

\begin{figure}[H]
    \centering
    \includegraphics[width=1\linewidth]{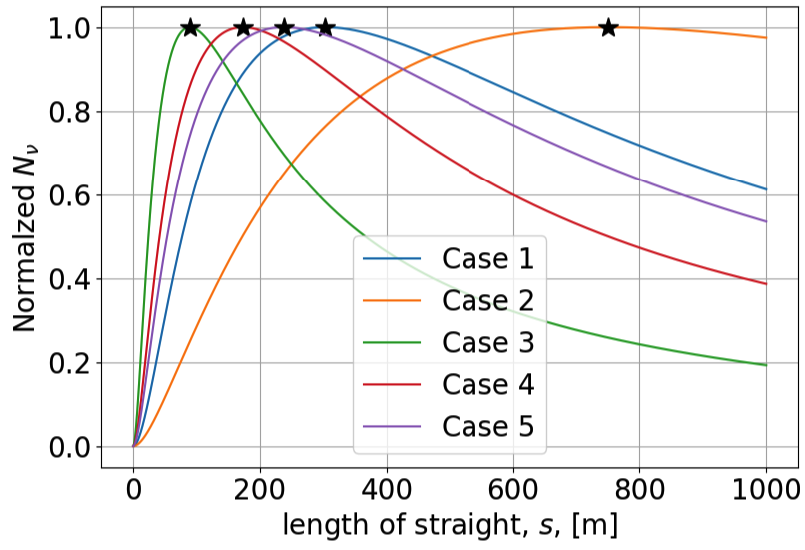}
    \caption{Plotting $N_\nu$ as a function of $s$ with Eq.~\ref{eq:N_nu} for different setups listed in Table~\ref{tab:valid_data_fermi} and \ref{tab:valid_data_ESS}, normalized with respect to the maximum $N_\nu$ for each case. The optimal values of $s$ are 89\,m, 173\,m, 238\,m, 303\,m and 750\,m, respectively, for the sets Case 1-5.}
    \label{fig:opt_s}
\end{figure}

Among the tunable parameters, the length $s$ is the most difficult parameter to fix, as it does not have a simple relation with the neutrino flux. Figure~\ref{fig:opt_s} shows an example where a fixed set of geometric parameters from Case 1~-~5 has been used to identify the optimum value of $s$. It demonstrates that each setup possesses a well-defined optimum: at first, the flux rises steeply with increasing $s$ since there is more time for the pions to decay in the straight, thus producing more muons and, subsequently, more neutrinos. When $s$ is extended further, the flux reaches a broad maximum, and then declines as the finite detector acceptance angle begins to dominate. 

For Case 1, the curve reaches a fairly broad maximum after 200\,m and peaks near 300\,m. A 180\,m straight therefore captures most of the neutrino flux while keeping the ring compact. One way of optimizing the flux for the combined Case 1+2 is to choose the value of $s$ which is at the intersection of the two curves, i.e. at around 450\,m, though this could be considered excessive from a cost point of view.

For Case~3, a low-energy ring for the ESSnuSB+ project, the curve peaks near 89\,m. An 81\,m straight thus offers reduced construction costs without significantly sacrificing the performance. 

Although the effect of other parameters on the neutrino flux is qualitatively clear, Eq.~\eqref{eq:N_nu} allows a quantitative comparison of their influence. By systematically varying the parameters, we can determine the optimal combination of detector radius, arc length, and detector distance that yields the highest neutrino flux while balancing other design constraints.

\begin{figure}[H]
    \centering
    \includegraphics[width=1\linewidth]{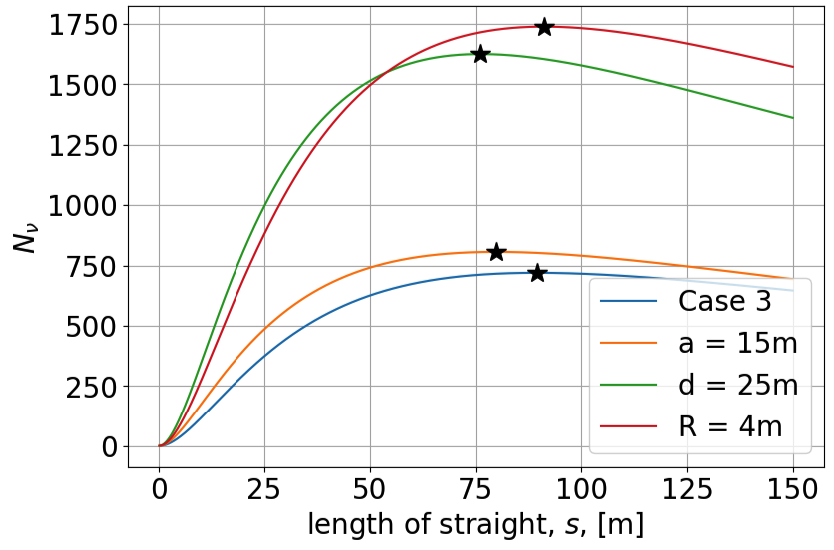}
    \caption{The flux fraction $N_\nu$ as a function of straight length $s$ for Case 3 and three parameter variations. The starting point, labeled 'Case 3', has the same parameter setup as Case 3 listed in Table~\ref{tab:valid_data_ESS}. In the other curves, a single parameter has been modified each time compared to Case 3, as marked in the label.}
    \label{fig:other_para}
\end{figure}

Figure~\ref{fig:other_para} shows an example where the initial parameters from Case 3 in Table~\ref{tab:valid_data_ESS} are modified one by one. By plotting $N_\nu(s)$ from Eq.~\eqref{eq:N_nu}, we can select a suitable length for the straight section for each configuration of the other facility parameters. In addition, we can conclude the contribution from each parameter to the flux and can use this to balance performance with cost. A sample code of this computation method is provided in Ref.~\cite{nuSTORM_opt}.

\section{Neutrino spectrum estimation}
In addition to the neutrino flux, we can also use Eq.~\eqref{eq:N_nu} to obtain the energy spectrum of the neutrinos arriving at the detector. This has previously been done by approximating the neutrino source to a point source~\cite{nuflux1997}, an approximation which is effective at higher energies, but becomes less accurate at lower energies and for short baselines. A more precise method of flux estimation is to apply the expression presented in the previous sections to a neutrino source which is distributed along the straight section, i.e. by considering the \mbox{nuSTORM} source as a line rather than a single point. 

At each point along the straight section, the distribution in energy and emission angle of the neutrinos in the muon rest frame is predicted by electroweak theory, and can be found, for example, in Ref.~\cite{Gaisser_Engel_Resconi_2016}. The distribution can be transformed into the corresponding distribution in emission angle and momentum in the lab frame, which is given by
\begin{equation}
    \frac{dn}{dE_\nu\,d\theta} =
        \begin{cases}
            \dfrac{f(z)\sin\theta}
            {E_\mu \left(1 - \beta \cos\theta\right)},
            & 0 \leq z \leq 1, \\
            0, & \text{otherwise}
        \end{cases}
        \label{eq:2dpdf}
\end{equation}
where 
\begin{equation}
    f(z) =
        \begin{cases}
            2z^2(3 - 2z), & \text{for } \nu_\mu \text{ or } \bar{\nu}_\mu  \\[6pt]
            12z^2(1 - z), & \text{for } \nu_e \text{ or } \bar{\nu}_e,
        \end{cases}
\end{equation}
and
\begin{equation}
    z = \frac{2 E_\nu E_{\mu} \left(1 - \beta \cos\theta\right)}{m_\mu^2}.
\end{equation}
Here, $E_\nu$ and $E_\mu$ are the neutrino and muon energies, respectively, in the lab frame. Details of this derivation are found in Appendix~\ref{app:nu_dis}. 

Figure~\ref{fig:2dpdf} shows an example of a $\nu_e$ distribution, calculated with Eq.~\eqref{eq:2dpdf} and assuming $P_\mu=600$\,MeV/c.
\begin{figure}[h]
    \centering
    \includegraphics[width=0.9\linewidth]{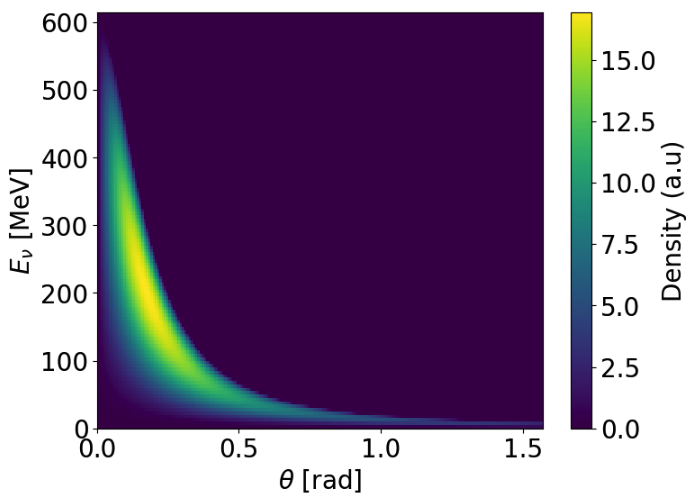}
    \caption{Two-dimensional probability density in neutrino energy $E_\nu$ and emission angle $\theta$ of electron neutrinos from the decay of 600\,MeV/c muons, computed with Eq.~\eqref{eq:2dpdf}.}
    \label{fig:2dpdf}
\end{figure}

Using Eq.~\eqref{eq:2dpdf}, the neutrino distribution can be obtained at any point $x$ along the straight section. For each point $x$, the two-dimensional distribution is integrated over the angle $\theta$ from 0 to $\theta_a$ defined with Eq.~\eqref{eq:theta_x}. This integration yields the neutrino distribution in energy, at the detector, collected from that point $x$. The distributions from all points along the straight section are then convoluted with the muon density in Eq.~\eqref{eq:mu_den} into the complete spectrum.

\begin{figure}[H]
    \centering
    \subfloat[Energy distribution of $\nu_e$ for Case 3.\label{fig:pdf_nu_e}]{
        \includegraphics[width=0.45\textwidth]{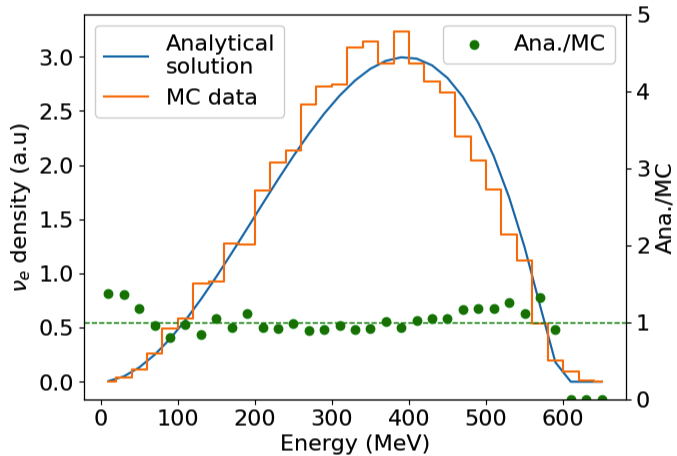}
    }
    \par\vspace{0.5cm}
    \subfloat[Energy distribution of $\nu_\mu$ for Case 3.\label{fig:pdf_nu_mu}]{
        \includegraphics[width=0.45\textwidth]{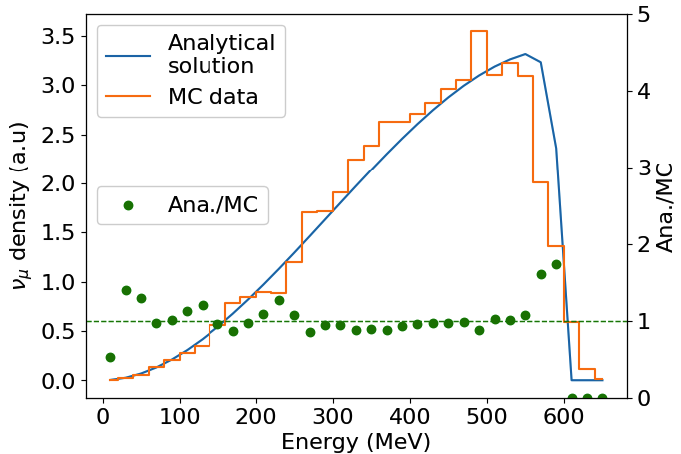}
    }

    \caption{The calculated and Monte Carlo (MC) energy distributions and their ratio for (a) $\nu_e$ and (b) $\nu_\mu$ in Case~3. The distributions have been normalized to their respective integrals. Refer to Table~\ref{tab:valid_data_ESS} for the facility parameters.}
    \label{fig:pdf_nu}
\end{figure}

Figures~\ref{fig:pdf_nu_e} and \ref{fig:pdf_nu_mu} show examples of the final electron and muon neutrino spectra at the detector for Case 3. For comparison, the equivalent spectra extracted from the Monte Carlo simulation are shown alongside the calculated values. In these examples, $R=2.5$\,m, $d=50$\,m and $s=81$\,m. 
Note that Eq.~\eqref{eq:2dpdf} assumes monochromatic muons, whereas the finite muon momentum acceptance is automatically included in the Monte Carlo simulations. Despite this discrepancy, the analytical distribution matches well with the Monte Carlo distribution. We have quantified the discrepancy between the two curves through the ratio between the Monte Carlo data and the analytical solution, which is shown as the green dots in Fig.~\ref{fig:pdf_nu}. The fact that the ratio is larger at the lower and higher energy ends is attributed to the fact that the Monte Carlo simulations include the muon momentum spread and therefore allow production of neutrinos up to 660\,MeV, whereas the calculation is limited to 610\,MeV. Furthermore, MC data are naturally low at the two ends.

By multiplying the neutrino energy density by the total number of neutrinos, which you can get from Eq.~\eqref{eq:N_nu}, we can predict the neutrino energy spectrum of a particular nuSTORM facility. A sample Python code to produce the neutrino energy density Fig.~\ref{fig:pdf_nu_e} and Fig.~\ref{fig:pdf_nu_mu} is provided in Ref.~\cite{PDF_nu_e}.
\section{Conclusion}
A nuSTORM facility produces neutrino beams of well-monitored intensity and with equal amounts of electron and muon flavor, from muons stored in a racetrack ring. The performance of the nuSTORM depends on the selection of several intertwined parameters. 

We have derived an analytical expression for the absolute number of neutrinos reaching the detector front face. The expression includes the pion and muon momenta and momentum acceptance, the lengths of the straight and the arc sections, and the detector radius and baseline. It is based on assumptions of the muon angular acceptance and uses a monochromatic pion pencil beam and muon pencil beam. Using the expression, it is possible to quickly calculate the number of neutrinos that are to be expected in the detector, before the racetrack ring has been fully designed, and without performing time-consuming Monte Carlo simulations. 

Five datasets from Monte Carlo simulations were used to validate the expression, which gave results of the same order of magnitude as the simulations for all of the five cases, which cover muon energies from hundreds of MeV to several GeV.

The analytical expression was used to determine the optimum length of the racetrack ring and to study the influence of other facility parameters on the neutrino flux. In addition, we complemented the expression with the neutrino energy probability density, thus yielding not only the total neutrino flux but also the neutrino energy spectrum seen at the detector. These results can be used to study the experiment performance at the conceptual stage of a nuSTORM project. Additional constraints from other considerations, e.g. limits on neutrino emission angle and energy from the detection perspective, can be easily applied alongside this expression to meet the requirements of a specific study.  

\section{Acknowledgment}
Funded by the European Union. Views and opinions expressed are, however, those of the author(s) only and do not necessarily reflect those of the European Union. Neither the European Union nor the granting authority can be held responsible for them.

This work was also supported by the Croatian Science Foundation under the project numbers HRZZ-DOK-NPOO-2023-10-1262 and HRZZ-DOK-2025-02-2372; within the framework of the second Swiss Contribution under grant MAPS IZ11Z0\_230193 by the Swiss National Science Foundation (SNSF) and the Croatian Science Foundation (HRZZ); Ministry of Science, Education and Youth of the Republic of Croatia via grant No. PK.1.1.10.0002.

The author thanks Dr Shinji Machida and Dr Jean-Baptiste Lagrange (Accelerator Science and Technology Centre (ASTeC), STFC) for stimulating discussions.

\newpage
\appendix
\section{Muon distribution from pion decay\label{app:uniform_Pmu}}
For simplicity, we use natural units and set the speed of light $c=1$ throughout the derivations in Appendices~\ref{app:uniform_Pmu}-\ref{app:nu_dis}.

In pion decay, the daughter muon energy $E^\prime$ and momentum $P^\prime$ in the pion rest frame, labelled with prime, are constants, related through
\begin{align}
    \begin{aligned} 
        P^\prime = \frac{m_{\pi}^2 - m_{\mu}^2 }{2m_{\pi}} 
        \quad \text{and} \quad
        E^\prime = \frac{ {m_{\pi}^2 + m_{\mu}^2} }{2m_{\pi}}. 
    \end{aligned}
    \label{eq:P'E'}
\end{align}
Here, we assume both $P^\prime$ and $E^\prime$ are in units of energy, with $m_\pi$ and $m_\mu$ being the mass of the pion and muon, respectively. 

The muon energy $E$ in the lab frame is given by the Lorentz boost from the pion rest frame:
\begin{equation}
    E  = \gamma(E^\prime + \beta P^\prime\cos\theta^\prime)
    \label{eq:mu_boost}
\end{equation}
where $\gamma$ is the pion relativistic Lorentz factor, $\beta$ is the pion relativistic speed, and $\theta^\prime$ is the muon emission angle in the pion rest frame. 

Since pions decay isotropically in their rest frame, which means uniformly in solid angle, the distribution of $\cos\theta^\prime$ is uniform. Since $P^\prime$ and $E^\prime$ are constant, Eq.~\eqref{eq:mu_boost} reveals that the energy of the muon has a unique relation with $\cos\theta^\prime$. The muon energy distribution, therefore, has the same uniform distribution as $\cos\theta^\prime$.

\section{Derivation of $P_{\pi}/P_{min} \approx 7/4$ \label{app:47ratio}}
We start from the assumption that the pion momentum $P_\pi$ is much larger than its mass $m_\pi$ and satisfies
\[
\beta = \frac{P_{\pi}}{E_{\pi}} \approx 1 \quad \text{when} \quad P_{\pi} \gg m_{\pi}\,,
\]
where, again, $\beta$ is the pion relativistic  speed. The momentum of the decayed muon in the lab frame transforms from the rest frame, labeled with a prime, as 
\begin{align} 
    \left\{ 
    \begin{aligned}    
        P_{x} &= P^\prime \sin{\theta^\prime} \\
        P_{z} &= \gamma (P^\prime \cos{\theta^\prime} + \beta E^\prime), \nonumber
    \end{aligned} \right.
\end{align}
where $\gamma$ is the pion Lorentz factor, and where the suffix $x$ and $z$ are the transverse and longitudinal directions. For the maximum and minimum muon momentum, $P_{\rm{max}}$ and $P_{\rm{min}}$, $\theta^\prime$ is 0 and 180 degrees, respectively. In both cases, the component $P_x\propto\sin\theta^\prime=0$, and we get
\begin{align}
    P_{\rm{max} }&=P_z(\theta=0) =\gamma(P^\prime + \beta E^\prime), \quad {\rm{and}} \label{eq:Pmax} \\
    P_{\rm{min} }&=P_z(\theta=180^\circ) =\gamma(-P^\prime + \beta E^\prime)\,. \nonumber 
\end{align}
The ratio between $P_{\rm{max}}$ and $P_{\min}$ is then
\begin{equation}
    \frac{P_{\max}}{P_{\min}} 
    = \frac{\gamma(P^\prime + \beta E^\prime)}{\gamma(-P^\prime + \beta E^\prime)}
    = 
    \frac{1 + \frac{P^\prime}{\beta E^\prime}}{1 - \frac{P^\prime}{\beta E^\prime}}. \nonumber
\end{equation}
The relation $P^\prime/(\beta E^\prime)$ can be extracted from two-body decay kinematics in Eq.~\eqref{eq:P'E'}:
\begin{equation}
    \frac{P^\prime}{\beta E^\prime} = 
    \frac{m_{\pi}^2 - m_{\mu}^2}{\beta (m_{\pi}^2 + m_{\mu}^2)} \nonumber
\end{equation}
In the limit when $\beta\rightarrow1$ we get
\begin{equation}
    \lim_{\beta \to 1}  \frac{P_{\max}}{P_{\min}} = 1.74493 \approx 1.75 =\frac{7}{4}.
    \nonumber
\end{equation}

On the other hand, with $E_\pi =\gamma m_\pi$, $P_{\rm{max}}$ in Eq.~\eqref{eq:Pmax} can be written as 
\begin{equation}
    P_{\mathrm{max}} = \frac{E_\pi}{m_\pi} \left(P^\prime+\beta E^\prime\right) ,
    \label{eq:P_max}
\end{equation}
which, by multiplying $P_\pi$ in both numerator and denominator, becomes
\begin{equation}
    P_{\mathrm{max}} = \frac{E_\pi}{m_\pi} \frac{P_\pi}{P_\pi}\left(P^\prime+\beta E^\prime\right) \nonumber\\
\end{equation}
Recognizing that $\beta=P_\pi/E_\pi$ gives 
\begin{equation}
    P_{\mathrm{max}} = \frac{P_\pi}{m_\pi} \left(\frac{P^\prime}{\beta}+E^\prime\right). \nonumber
\end{equation}
In the limit of $\beta\rightarrow1$, $(P^\prime/\beta+E^\prime)\rightarrow m_\pi$ and therefore $P_{\mathrm{max}}\approx P_\pi$. As a result,
\begin{equation}
    \frac{P_{\max}}{P_{\min}} \approx \frac{P_{\pi}}{P_{\min}} \approx \frac{7}{4}.
\end{equation}

\section{Derivation of the neutrino angular distribution in the lab frame\label{app:headlight}}
This derivation is mainly inspired by Andrew's book \cite{headlight_src}, chapter 3.7.

Suppose there are some muons with the same energy and momentum in their rest frame, again labeled with a prime. Those muons decay and give out $N$ neutrinos per type. In the muon rest frame, the muons decay isotropically, and the distribution is constant everywhere in solid angle $\Omega^{\prime}$:
\begin{equation}
\label{eq:dNdOmega}
        \frac{d N}{d \Omega^{\prime}} = \text{constant} \nonumber
\end{equation}
which can be expanded to
\begin{align}
    \frac{d N}{\sin\theta^\prime \,d\theta^\prime d \phi^{\prime}} = C\,.\label{eq:dNdthetadphi}
\end{align}
Integrating over the azimuthal angle $\phi'$, we reach
\begin{align}
    \frac{d N}{d \theta^\prime} =2 \pi C \sin \theta^\prime\,, \label{eq:dNdtheta}
\end{align}
and by integrating Eq.~\eqref{eq:dNdthetadphi} over all solid angles, we get $C=N/4\pi$.

In the lab frame, the angular distribution of the neutrinos can be written as: 
\begin{equation}
    \rho(\theta)=\frac{d N}{d \theta}=\frac{d N}{d \theta^\prime} \frac{d \theta^\prime}{d \theta} \nonumber
\end{equation}
where $dN/d\theta^\prime$ is known from Eq.~\eqref{eq:dNdtheta}. The second factor, $d\theta^\prime/d\theta$, is given by the relativistic kinematics. 

The energy $E$ and forward momentum $P_z$ of a neutrino transform from the rest frame to the lab frame according to 
\begin{equation}
    \left\{    
    \begin{aligned}
        E &= \gamma\left(E^\prime+\beta P_z^{\prime}\right) \\ 
        P_z &= \gamma\left(P_z^{\prime}+\beta E^\prime\right),
     \label{eq:E_Pz}
    \end{aligned}\right.
\end{equation}
where $\gamma$ is the muon relativistic Lorentz factor and $\beta$ is the muon relativistic velocity.

For a massless daughter, $E^{(\prime)} = P^{(\prime)}$ in either reference frame. With $P^{(\prime)}_z=P^{(\prime)} \cos \theta^{(\prime)}$, Eq.~\eqref{eq:E_Pz} becomes
\begin{empheq}
    [left=\empheqlbrace]{align}
    P &= \gamma\left(P^\prime+ \beta P^\prime\cos\theta^\prime\right) \label{eq:top} \\
    P\cos\theta &= \gamma\left(P^\prime\cos\theta^\prime + \beta P^\prime\right).
    \label{eq:bottom} 
\end{empheq}
Dividing Eq.~\eqref{eq:bottom} by Eq.~\eqref{eq:top} gives
\begin{align}
  \cos \theta=\frac{\cos \theta^\prime+\beta}{1+\beta \cos \theta^\prime}, \nonumber
\end{align}
which, by making $\cos\theta^\prime$ the subject, can be arranged to
\begin{align}
  \cos \theta^\prime=\frac{\cos \theta-\beta}{1-\beta \cos \theta}.
  \label{eq:dcos_theta}
\end{align}
Now, we differentiate both sides with respect to $\cos\theta$:
\begin{align}
    \frac{d\cos\theta^\prime}{d \cos \theta} &=\frac{1-\beta^2}{(1-\beta \cos \theta)^2}  \nonumber  \\
    \Longrightarrow \qquad
    \frac{\sin \theta^\prime \,d\theta^\prime}{\sin \theta \,d \theta} &=\frac{1-\beta^2}{(1-\beta \cos \theta)^2} \nonumber \\
    \Longrightarrow \qquad \qquad  \,\,
    \frac{d \theta^\prime}{d \theta}&=\frac{\sin \theta}{\sin \theta^\prime} \frac{1-\beta^2}{(1-\beta \cos \theta)^2} \nonumber
\end{align}

The neutrino angular distribution in the lab frame then becomes
\begin{align}
    \rho(\theta)&=\frac{d N}{d \theta}=\frac{d N}{d \theta^\prime} \frac{d \theta^\prime}{d \theta} \nonumber\\
    \Longrightarrow \qquad
    &=2 \pi C \sin \theta^\prime \frac{\sin \theta}{\sin \theta^\prime} \frac{1-\beta^2}{(1-\beta \cos \theta)^2} \nonumber\\
    \Longrightarrow \qquad
    &=\frac{N}{2} \sin\theta \frac{1-\beta^2}{(1-\beta \cos \theta)^2}\,. \nonumber
\end{align}
Finally, the angular probability density can be written as 
\begin{align}
    \eta(\theta)=\frac{\rho(\theta)}{N}=\frac{1}{2} \sin\theta \frac{1-\beta^2}{(1-\beta \cos \theta)^2}\,, \nonumber
\end{align}
which, as required, satisfies 
\begin{equation}
    \int_{0}^{\pi} \eta(\theta) \, d\theta=1. \nonumber
\end{equation}

\section{Neutrino angular and energy distribution in the lab frame \label{app:nu_dis}}

The differential neutrino distribution in energy and polar angle in the muon rest frame is~\cite{Gaisser_Engel_Resconi_2016}
\begin{equation}
    \frac{dn}{dy^{\prime}\,d\cos\theta^\prime} =
        \begin{cases}
             f(y^\prime)/2,
            & 0 \leq y^\prime \leq 1, \\
            0, & \text{otherwise}
        \end{cases}
        \label{eq:nu_rest_dis}
\end{equation}
where
\begin{equation}
    f(y^\prime) =
        \begin{cases}
            2y^{\prime2}(3 - 2y^\prime), & \text{for } \nu_\mu \text{ or } \bar{\nu}_\mu  \\[6pt]
            12y^{\prime2}(1 - y^\prime), & \text{for } \nu_e \text{ or } \bar{\nu}_e,
        \end{cases}
        \nonumber
\end{equation}
 and $y^{\prime}$ is defined as 
\begin{equation}
    y^{\prime} = \frac{2E_\nu^{\prime}}{m_\mu} \,,
    \label{eq:y'}
\end{equation}
with $m_\mu$ being the muon rest mass. Here, $E_\nu'$ is the energy of the neutrino in the muon rest frame, which can be expressed in terms of the muon Lorentz factor $\gamma$, the relativistic muon speed $\beta$ and the neutrino energy $E_\nu$ and emission angle $\theta$ in the lab frame:
\begin{equation}
    E_\nu' = \gamma\,E_\nu\bigl(1-\beta\cos\theta\bigr), \nonumber
\end{equation}
With this, Eq.~\eqref{eq:y'} becomes
\begin{equation}
    y^\prime = \frac{2}{m_\mu}\gamma E_\nu\bigl(1-\beta\cos\theta\bigr). 
    \label{eq:y'2}
\end{equation}
The distribution can be transformed to the lab frame through
\begin{equation}
    \frac{dn}{dE_\nu\,d\theta} =
    \frac{dn}{dy'\,d\cos\theta'} \left|\det\mathbf{J}\right|, \label{eq:dndE_transform}
\end{equation}
where $\det\mathbf{J}$ denotes the determinant of the Jacobian matrix of the transformation from ($E_\nu,\, \theta)$ to $(y^\prime, \, \cos\theta^\prime)$:
\begin{equation}
    \det \mathbf{J} =
     \frac{\partial y'}{\partial E_\nu} \frac{\partial \cos\theta'}{\partial \theta}
    -\frac{\partial y'}{\partial \theta} \frac{\partial \cos\theta'}{\partial E_\nu}.
    \label{eq:detJ}
\end{equation}
By differentiating Eq.~\eqref{eq:y'2}, we get
\begin{equation}
    \frac{\partial y'}{\partial E_\nu}
    =
    \frac{2\gamma}{m_\mu}
    \left(1-\beta\cos\theta\right).
    \label{eq:dy'dE}
\end{equation}

Now, we differentiate Eq.~\eqref{eq:dcos_theta} with respect to $E_\nu$ and $\theta$, respectively, and obtain the following expressions:
\begin{equation}
    \frac{\partial \cos\theta'}{\partial E_\nu}=0 \quad \text{and}  \quad
    \frac{\partial \cos\theta'}{\partial \theta}
    =
    -\frac{\sin\theta}
    {\gamma^2\left(1-\beta\cos\theta\right)^2}.
    \label{eq:dtheta'dE}
\end{equation}

These, we insert in Eq.~\eqref{eq:detJ}, which then becomes
\begin{align}
   |\det\mathbf{J}| = \frac{2\sin\theta}{E_\mu\left(1-\beta\cos\theta\right)}\,.
\end{align}
\medskip

By combining the left-hand side of Eq.~\eqref{eq:nu_rest_dis} with Eqs.~\eqref{eq:dndE_transform} and \eqref{eq:nu_lab_dis} we now reach 
\begin{align}
    \frac{dn}{dy^{\prime}\,d\cos\theta^\prime} &=\frac{dn}{dE_\nu\,d\theta} \frac{1}{|\det\mathbf{J}|} \nonumber \\
    &=\frac{dn}{dE_\nu\,d\theta}\frac{E_\mu\left(1-\beta\cos\theta\right)}{2\sin\theta} \,. 
    \label{eq:nu_lab_dis}
\end{align}
The right-hand side of Eq.~\eqref{eq:nu_rest_dis} becomes  
\begin{align}
    \frac{f(y^\prime)}{2} = \frac{f\bigl(2\gamma E_\nu(1-\beta\cos\theta)/m_\mu\bigr)}{2}\,. \nonumber
\end{align}
We now define a new variable $z = 2\gamma E_\nu\bigl(1-\beta\cos\theta\bigr)/m_\mu$ so that Eq.~\eqref{eq:nu_lab_dis}, can be expressed as 
\begin{equation}
    \frac{dn}{dy^{\prime}\,d\cos\theta^\prime} 
    =\frac{dn}{dE_\nu\,d\theta}\frac{E_\mu\left(1-\beta\cos\theta\right)}{2\sin\theta} 
    = \frac{f(z)}{2}  \nonumber
\end{equation}
The neutrino distribution in energy and angle, in the lab frame, is lastly given by
\begin{equation}
    \frac{dn}{dE_\nu\,d\theta} = 
        \begin{cases}
        \frac{ f(z)\sin\theta}{E_\mu\left(1-\beta\cos\theta\right)}  ,
        & 0 \leq z \leq 1, \\
        0, & \text{otherwise}\,.
        \end{cases}
    \nonumber
\end{equation}

\bibliography{apssamp}

@techreport{Geer2011,
    author      = {Geer, Steve},
    title       = {How Many Muons Do We Need to Store in a Ring
                   for Neutrino Cross-Section Measurements?},
    institution = {Fermi National Accelerator Laboratory},
    number      = {FERMILAB-FN-0924-APC},
    year        = {2011},
    month       = jul,
    doi         = {10.2172/1022119},
    eprint      = {1204.6234},
    archivePrefix = {arXiv},
    primaryClass = {physics.acc-ph}
}

@phdthesis{fermi_nu,
title = {Design and simulation of the nuSTORM facility},
author = {Liu, Ao},
school = {Indiana University},
doi = {10.2172/1209538},
url = {https://www.osti.gov/biblio/1209538}, 
year = {2015},
month = {apr}
}

@article{essnusb2023,
   title={The ESSnuSB Design Study: Overview and Future Prospects},
   volume={9},
   ISSN={2218-1997},
   url={http://dx.doi.org/10.3390/universe9080347},
   DOI={10.3390/universe9080347},
   number={8},
   journal={Universe},
   publisher={MDPI AG},
   author={Alekou, A. and others},
   collaboration = {The ESSnuSB Collaboration},
   year={2023},
   month=jul, pages={347} }

@article{ESSnuSB_CDR,
	doi = {10.1140/epjs/s11734-022-00664-w},
  
	url = {https://doi.org/10.1140%2Fepjs%2Fs11734-022-00664-w},
  
	year = 2022,
	month = {11},
  
	publisher = {Springer Science and Business Media {LLC}},
  
	volume = {231},
  
	number = {21},
  
	pages = {3779--3955},
  
	author = {Alekou, A.  and others},
    collaboration = {ESSnuSB Collaboration},
	title = {The European Spallation Source neutrino super-beam conceptual design report},
  
	journal = {The European Physical Journal Special Topics}
}

@misc{nustorm2022,
      title={Neutrinos from Stored Muons (nuSTORM)}, 
      author={L. Alvarez Ruso and T. Alves and S. Boyd and A. Bross and P. R. Hobson and P. Kyberd and J. B. Lagrange and K. Long and X. -G. Lu and J. Pasternak and M. Pfaff and C. Rogers},
      year={2022},
      eprint={2203.07545},
      archivePrefix={arXiv},
      url={https://arxiv.org/abs/2203.07545}, 
      primaryClass={hep-ex}
}

@techreport{nustorm_cern,
      author        = "Ahdida, C.C. and Appleby, R. and Bartmann, W. and Bauche,
                       J. and Calviani, M. and Gall, J. and Gilardoni, S. and
                       Goddard, B. and Hessler, C. and Huber, P. and
                       Efthymiopoulos, I. and Lagrange, J.B. and Lamont, M. and
                       Long, K. and Osborne, J.A. and Pasternak, J. and Soler,
                       F.J.P. and Tygier, S. and Velotti, F.M.",
      title         = "{nuSTORM at CERN: Feasibility Study}",
      institution   = "CERN",
      reportNumber  = "CERN-PBC-REPORT-2019-003",
      address       = "Geneva",
      year          = "2020",
      url           = "https://cds.cern.ch/record/2654649",
      doi           = "10.17181/CERN.FQTB.O8QN",
}

@misc{nustorm1980,
author = {David Neuffer},
title = {Design considerations for a muon storage ring},
url = {https://inspirehep.net/files/012057007c5746a3feb1972a97f06ab0},
year = {1980},
note = {(Accessed on 03/31/2024)}
}

@book{headlight_src,
  author    = {Andrew M. Steane},
  title     = {Relativity Made Relatively Easy: Volume 1},
  year      = {2012},
  publisher = {Oxford University Press},
  address   = {Oxford},
  volume    = {1},
  isbn      = {9780199662869}
}

@article{nuflux1997,
  title = {Neutrino beams from muon storage rings: Characteristics and physics potential},
  author = {Geer, S.},
  journal = {Phys. Rev. D},
  volume = {57},
  issue = {11},
  pages = {6989--6997},
  numpages = {0},
  year = {1998},
  month = {Jun},
  publisher = {American Physical Society},
  doi = {10.1103/PhysRevD.57.6989},
  url = {https://link.aps.org/doi/10.1103/PhysRevD.57.6989}
}

@book{Gaisser_Engel_Resconi_2016, 
    place={Cambridge}, 
    edition={2},
    title={Cosmic Rays and Particle Physics}, 
    publisher={Cambridge University Press}, 
    author={Gaisser, Thomas K. and Engel, Ralph and Resconi, Elisa}, 
    year={2016}
}

@article{nu_fermi_far,
   title={Light sterile neutrino sensitivity at the nuSTORM facility},
   volume={89},
   ISSN={1550-2368},
   url={http://dx.doi.org/10.1103/PhysRevD.89.071301},
   DOI={10.1103/physrevd.89.071301},
   number={7},
   journal={Physical Review D},
   publisher={American Physical Society (APS)},
   author={Adey, D. and others},
   collaboration = {The nuSTORM Collaboration},
   year={2014},
   month=Apr }

@misc{xsuite,
  title = {Xsuite: an integrated beam physics simulation framework},
  author = {G. Iadarola and R. De Maria and S. Lopaciuk and A. Abramov and X. Buffat and D. Demetriadou and L. Deniau and P. Hermes and P. Kicsiny and P. Kruyt and A. Latina and L. Mether and K. Paraschou and Sterbini and F. Van Der Veken and P. Belanger and P. Niedermayer and D. Di Croce and T. Pieloni and L. Van Riesen-Haupt},
  year = {2023},
  eprint = {2310.00317},
  archiveprefix = {arXiv},
  primaryclass = {physics.acc-ph},
  url = {https://arxiv.org/abs/2310.00317},
}

@article{BDSIM,
  title = {BDSIM: An accelerator tracking code with particle–matter interactions},
  volume = {252},
  issn = {0010-4655},
  url = {http://dx.doi.org/10.1016/j.cpc.2020.107200},
  doi = {10.1016/j.cpc.2020.107200},
  journal = {Computer Physics Communications},
  publisher = {Elsevier BV},
  author = {Nevay, L.J. and Boogert, S.T. and Snuverink, J. and Abramov, A. and Deacon, L.C. and Garcia-Morales, H. and Lefebvre, H. and Gibson, S.M. and Kwee-Hinzmann, R. and Shields, W. and Walker, S.D.},
  year = {2020},
  month = jul,
  pages = {107200},
}

@misc{PDF_nu_e,
  author       = {T. W. Choi},
  title        = {{PDF\_nu\_e}: Jupyter notebook for LEnuSTORM},
  howpublished = {\url{https://github.com/tingchoi518/LEnuSTORM/blob/main/PDF_nu_e.ipynb}},
  note         = {GitHub repository, accessed 4 September 2026},
  year         = {2026}
}

@misc{nuSTORM_opt,
  author       = {T. W. Choi},
  title        = {Optimizing nuSTORM with an anlystical expression : Jupyter notebook for LEnuSTORM},
  howpublished = {\url{https://github.com/tingchoi518/LEnuSTORM/blob/main/nuSTORM_formula.ipynb}},
  note         = {GitHub repository, accessed 4 September 2026},
  year         = {2026}
}

@inproceedings{choi:ipac2026-mop1098,
  author = {T. W. Choi and M. Olveg{\aa}rd and M. Topp-Mugglestone},
  title = {{LEnuSTORM}: A Low-Energy Muon Storage Ring for Neutrino Cross-Section Measurements},
  booktitle = {Proc. IPAC'26},
  pages = {321--324},
  paper = {MOP1098},
  venue = {Deauville, France},
  series = {International Particle Accelerator Conference},
  number = {17},
  publisher = {JACoW Publishing, Geneva, Switzerland},
  year = {2026},
  issn = {2673-5350},
  isbn = {978-3-95450-252-3},
  doi = {10.18429/JACoW-IPAC2026-MOP1098},
}

\end{document}